\documentclass[twocolumn,preprintnumbers,amsmath,amssymb,prb]{revtex4}

\usepackage{graphicx}
\usepackage{dcolumn}
\usepackage{bm}
\usepackage{amssymb}
\usepackage{epstopdf}
\usepackage{color}

\begin{document}

\title{Floquet engineering of excitons in semiconductor quantum dots}

\author{I. V. Iorsh$^{1,2}$}
\author{D. A. Zezyulin$^{1}$}
\author{S. A. Kolodny$^{1,2}$}
\author{R. E. Sinitskiy$^{2}$}
\author{O. V. Kibis$^{2,}$}\email{Oleg.Kibis(c)nstu.ru}
\affiliation{$^1$Department of Physics and Engineering, ITMO~ University, Saint-Petersburg 197101, Russia}
\affiliation {$^2$Department of Applied and Theoretical Physics, Novosibirsk~State~Technical~University,
Karl~Marx~Avenue~20,~Novosibirsk~630073,~Russia}

\begin{abstract}
Within the Floquet theory of periodically driven quantum systems, we demonstrate that a high-frequency  electromagnetic field can be used as an effective tool to control excitonic properties of semiconductor quantum dots (QDs). It is shown, particularly, that the field both decreases the exciton binding energy and dynamically stabilizes the exciton, increasing its radiative lifetime. The developed theory can serve as a basis for the ultrafast method to tune spectral characteristics of the QD-based photon emitters by a high-frequency field.
\end{abstract}
\maketitle

\section{Introduction}
Controlling electronic properties of quantum materials by an off-resonant high-frequency electromagnetic field, which is based on the Floquet theory of periodically driven quantum systems (Floquet engineering), has become an established research area of modern physics~\cite{Oka_2019,Basov_2017,Eckardt_2015,Goldman_2014,Bukov_2015,Casas_2001,Kibis_2020_1}. Since frequency of the off-resonant field is assumed to be far from characteristic
resonant frequencies of the electronic system, it cannot be absorbed by electrons and only
``dresses'' them (the dressing field). Therefore, the effect of such a dressing field is purely in renormalizing parameters of the electronic Hamiltonian. As a result, the dressing field can crucially modify physical
properties of various condensed-matter nanostructures, including semiconductor quantum wells~\cite{Lindner_2011,Pervishko_2015,Dini_2016}, quantum rings~\cite{Kibis_2011,Koshelev_2015,Kibis_2015,Kozin_2018}, topological insulators~\cite{Kozin_2018_1,Rechtsman_2013,Wang_2013,Torres_2014,Calvo_2015,Mikami_2016}, carbon nanotubes~\cite{Kibis_2021_1},
graphene and related two-dimensional materials~\cite{Oka_2009,Kibis_2010,Iurov_2017,Iurov_2013,Syzranov_2013,Usaj_2014,Perez_2014,Glazov_2014,Sentef_2015,Sie_2015,Kibis_2017,Iorsh_2017,Iurov_2019,Iurov_2020,Cavalleri_2020}, etc.

Among the most actively studied nanostructures, semiconductor quantum dots (QDs) take deserved place since they are the only stable source of single photons required for quantum communications and quantum metrology~\cite{somaschi2016near,ding2016demand,arakawa2020progress,muller2017quantum,bennett2016cavity}.
As a consequence, the QDs are considered as indispensable building blocks of modern quantum technology. A common problem in QD-based single-photon emitters is the control over their spectral characteristics. While the central frequency and linewidth of the photon emission are defined by the QD material and geometry and thus are fixed by the QD fabrication protocol, it is required for many applications in optical networks to tune the QD spectral characteristics dynamically. Since the photon emission in QDs originates from the recombination of electron-hole pairs (excitons), the optical properties of the QDs are totally dominated by the excitonic response. It should be noted that this response is clearly pronounced in QDs due to the large excitonic binding energies and oscillator strengths arisen from the strong quantum confinement of excitons~\cite{glutsch2004excitons}. Currently, the conventional way to achieve the spectral tunability of QDs is the gate voltage which allows to tune the exciton energy and oscillator strength by the electrostatic potential via the Stark effect~\cite{hallett2018electrical, trivedi2020generation}. In the present article, we will develop theoretically the alternative way to tune the exciton parameters by a dressing electromagnetic field.

The article is organised as follows. In Sec.~II, we construct the effective Hamiltonian describing an exciton in a semiconductor QD driven by a high-frequency off-resonant electromagnetic field. In Sec.~III, the Floquet problem with the effective Hamiltonian is solved and the found solutions of the problem are analyzed. The two last sections contain conclusion and acknowledgments.

\section{Model}
\begin{figure}[h!]
\centering\includegraphics[width=1.0\columnwidth]{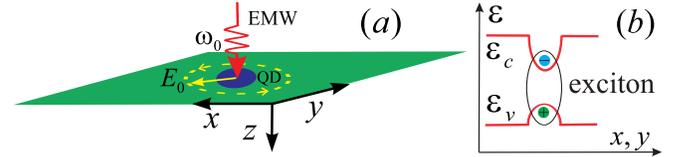}
\caption{Sketch of the system under consideration: (a) The quantum dot (QD) with parabolic confinement potential irradiated by the circularly polarized electromagnetic wave (EMW) with the frequency $\omega_0$ and the electric field amplitude $E_0$; (b) The energy band structure of the parabolic QD containing the electron-hole pair (exciton), where $\varepsilon_{c(v)}$ is the band edge of the conduction (valence) band of the semiconductor.} \label{fig:prelim}
\end{figure}
Let us consider a QD with the parabolic potential confining motion of electron and hole in the $x,y$ plane (parabolic QD), assuming that its in-plane size, $L$, much exceeds its size along the $z$ axis (see Fig.~1). Then the exciton Hamiltonian is~\cite{que1992excitons}
\begin{equation}\label{H}
\hat{\cal H}_{{X0}}=\sum_{j=e,h}\frac{\hat{\mathbf{p}}_j^2}{2m_j}+U_\Omega(\mathbf{r}_e,\mathbf{r}_h)+U_C(\mathbf{r}_e,\mathbf{r}_h),
\end{equation}
where $\hat{p}_{e(h)}$ is the electron (hole) momentum operator, $m_{e(h)}$ is the electron (hole) effective mass, $\mathbf{r}_{e(h)}$ is the in-plane radius-vector of electron (hole),
\begin{equation}\label{UOm}
U_\Omega(\mathbf{r}_e,\mathbf{r}_h)=\sum_{j=e,h}\frac{m_j\Omega^2\mathbf{r}_j^2}{2},
\end{equation}
is the potential energy of electron and hole in the parabolic confining potential, $\Omega$ is the frequency of electron and hole oscillations in this potential,
\begin{equation}\label{UC}
U_C(\mathbf{r}_e,\mathbf{r}_h)=-\frac{e^2}{\epsilon |\mathbf{r}_e-\mathbf{r}_h|},
\end{equation}
is the potential energy of the Coulomb interaction between electron and hole, and
$\epsilon$ is the permittivity. Let us irradiate the QD by a circularly polarized electromagnetic wave (EMW) with the frequency $\omega_0$ and the electric field amplitude $E_0$, which propagates along the $z$ axis (see Fig.~1a). Describing the interaction between the exciton and the EMW field in the Gaussian system of units within the conventional minimal coupling scheme, the exciton Hamiltonian (\ref{H}) in the presence of the EMW reads
\begin{equation}\label{HEMW}
\hat{\cal H}_{X}=\sum_{j=e,h}\frac{[\hat{\mathbf{p}}_j-e\xi_j\mathbf{A}(t)/c]^2}{2m_j}+U_\Omega(\mathbf{r}_e,\mathbf{r}_h)+U_C(\mathbf{r}_e,\mathbf{r}_h),
\end{equation}
where $e=-|e|$ is the elementary electron charge defined as a negative quantity, $\xi_e=1$, $\xi_h=-1$, and
\begin{equation}\label{A}
\mathbf{A}(t)=(A_x,A_y)=[cE_0/\omega_0](\sin\omega_0 t,\cos\omega_0 t)
\end{equation}
is the vector potential of the EMW (the Weyl gauge with the zero scalar potential is used). The Schr\"odinger equation with the periodically time-dependent Hamiltonian (\ref{HEMW}) describes the Floquet problem for the exciton dressed by the field (\ref{A}), which will be under consideration in the following. To simplify the problem, let us transform the Hamiltonian (\ref{HEMW}) with the unitary transformation
\begin{equation}\label{K}
\hat{\mathcal{\cal U}}(t)=\exp\hspace{-0.2em}{\left[\frac{i}{\hbar c}\sum_{j=e,h}\frac{e}{m_j}\int^t \hspace{-0.5em} dt' \left( \mathbf{A}(t')\xi_j\hat{\mathbf{p}}_j-\frac{eA^2(t')}{2c}\right)\right]},
\end{equation}
which is the Kramers-Henneberger  transformation~\cite{Kramers,henneberger1968perturbation} generalized to the considered case of electron-hole pair. Then the transformed Hamiltonian (\ref{HEMW}) reads
\begin{align}\label{Ht}
&\hat{\cal H}=\hat{\cal U}^\dagger(t)\hat{\cal H}_X\hat{\cal U}(t) -
i\hbar\hat{\cal U}^\dagger(t)\partial_t
\hat{\cal U}(t)=\sum_{j=e,h}\frac{\hat{\mathbf{p}}_j^2}{2m_j}\,+\nonumber\\
&U_\Omega(\mathbf{r}_e-\mathbf{r}_{e}^\prime(t),\mathbf{r}_h-\mathbf{r}_{h}^\prime(t))+U_C(\mathbf{r}_e-\mathbf{r}_{e}^\prime(t),\mathbf{r}_h-\mathbf{r}_{h}^\prime(t)),
\end{align}
where
\begin{equation}\label{rR}
\mathbf{r}_{e(h)}^\prime(t)=\xi_{e(h)}\bar{r}_{e(h)}(\cos\omega_0 t, -\sin\omega_0 t)
\end{equation}
is the radius-vector describing the classical circular trajectory of electron (hole) in the field (\ref{A}), and $\bar{r}_{e(h)}=|e|E_0/m_{e(h)}\omega_0^2$ is the radius of the trajectory. It follows from Eq.~(\ref{Ht}) that the unitary transformation (\ref{K}) removes the coupling of the momentums $\hat{\mathbf{p}}_{e,h}$ to the vector potential $\mathbf{A}(t)$ in the Hamiltonian and transfers the time dependence from the kinetic energy of electron and hole to their potential energies (\ref{UOm}) and (\ref{UC}), shifting the electron and hole coordinates $\mathbf{r}_{e,h}$ by the radius-vectors (\ref{rR}).

The Hamiltonian (\ref{Ht}) is still physically equal to the exact Hamiltonian of irradiated exciton (\ref{HEMW}). To proceed within the conventional Floquet theory, one can apply the $1/\omega_0$ expansion (the Floquet-Magnus expansion~\cite{Eckardt_2015,Goldman_2014,Bukov_2015,Casas_2001}) in order to turn the periodically time-dependent Hamiltonian (\ref{Ht}) into the effective stationary Hamiltonian with the main term
\begin{equation}\label{Hef}
\hat{\cal H}_{\mathrm{eff}}=\hat{\cal H}_{0},
\end{equation}
where $\hat{\cal H}_{0}$ is the zero harmonic of the Fourier expansion of the Hamiltonian (\ref{Ht}), $\hat{\cal H}=\sum_{n=-\infty}^{\infty}\hat{\cal H}_ne^{in\omega_0 t}$ (see Appendix for details).  Omitting the coordinate-independent term
$$V_0=\frac{(eE_0\Omega)^2}{2\mu\omega_0^4},$$
the effective Hamiltonian (\ref{Hef}) reads
\begin{equation}\label{Heff}
\hat{\cal
H}_{\mathrm{eff}}=\sum_{j=e,h}\left[\frac{\hat{\mathbf{p}}_j^2}{2m_j}+\sum_{j=e,h}\frac{m_j\Omega^2\mathbf{r}_j^2}{2}\right]
+U_0(\mathbf{r}_e,\mathbf{r}_h),
\end{equation}
where the potential
\begin{equation}\label{U0}
U_0(\mathbf{r}_e,\mathbf{r}_h)=\frac{1}{2\pi}\int_{-\pi}^{\pi}U_C(\mathbf{r}_e-\mathbf{r}_{e}^\prime(t),\mathbf{r}_h-\mathbf{r}_{h}^\prime(t))d(\omega
t)
\end{equation}
should be treated as the Coulomb potential renormalized by the dressing field (\ref{A}) (the dressed Coulomb potential) which turns into the ``bare'' Coulomb potential (\ref{UC}) in the absence of the dressing field ($E_0=0$). Substituting Eqs.~(\ref{UC}) and (\ref{rR}) into Eq.~(\ref{U0}), the effective Hamiltonian (\ref{Heff}) can be rewritten  as
\begin{equation}\label{Hefff}
\hat{\cal H}_{\mathrm{eff}}=\frac{\hat{\mathbf{P}}^2}{2M}+\frac{M\Omega^2\mathbf{R}^2}{2}+\frac{\hat{\mathbf{p}}^2}{2\mu}+\frac{\mu\Omega^2{r}^2}{2}+U_0({r}),
\end{equation}
where $\mathbf{R}=(m_e\mathbf{r}_e+m_h\mathbf{r}_h)/(m_e+m_h)$ is the radius-vector of the exciton center of mass,  ${\mathbf{r}}=\mathbf{r}_e-\mathbf{r}_h$ is the radius-vector of relative position of electron and hole, $M=m_e+m_h$ is the total exciton effective mass, $\mu=m_em_h/(m_e+m_h)$ is the reduced exciton mass, $\hat{\mathbf{P}}=-i\hbar\nabla_{\mathbf{R}}$ is the operator of center mass momentum, $\hat{\mathbf{p}}=-i\hbar\nabla_{\mathbf{r}}$ is the operator of momentum of relative motion of electron-hole pair,
\begin{eqnarray}\label{U00}
U_0({r})&=&
\left\{\begin{array}{rl}
-({2e^2}/{\pi r_0\epsilon})K\left({r}/{r_0}\right),
&{r}/{r_0}\leq1\\\\
-({2e^2}/{\pi r\epsilon})K\left({r_0}/{r}\right),
&{r}/{r_0}\geq1
\end{array}\right.
\end{eqnarray}
is the dressed Coulomb potential (\ref{U0}) written explicitly, the function ${K}(\zeta)$ is the elliptic  integral of the first kind, and
\begin{equation}\label{r0}
r_0=\frac{|e|E_0}{\mu\omega_0^2}
\end{equation}
is the sum of the radiuses of the electron and hole circular trajectories (\ref{rR}). Thus, the exact time-dependent Hamiltonian (\ref{HEMW}) in the high-frequency limit reduces to the approximate stationary Hamiltonian (\ref{Hefff}) with the dressed potential (\ref{U00}). As expected, the effective Hamiltonian (\ref{Hefff}) turns into the exact exciton Hamiltonian (\ref{H}) if the dressing field (\ref{A}) is absent ($E_0=0)$.

\section{Results and discussion}
Since the Hamiltonian (\ref{Hefff}) allows for the separation of the variables $\mathbf{R}$ and $\mathbf{r}$, its eigenfunctions $\Psi$ can be factorized, $\Psi=\Psi_1(\mathbf{R})\Psi_2(\mathbf{r})$. It follows from the Hamiltonian that $\Psi_1(\mathbf{R})$ is the well-known eigenfunction of the quantum harmonic oscillator with the eigenfrequency $\Omega$ and the mass $M$. Since $\mathbf{R}$-dependent part of the Hamiltonian is unaffected by the dressing field (\ref{A}), in what follows we assume that an irradiated exciton remains in the ground state of the oscillator with the energy $\hbar\Omega/2$. Since the dressed Coulomb potential (\ref{U00}) keeps the axial symmetry of the exciton, the
$z$-component of the angular momentum of relative exciton motion, $m$, is the conserved quantum number. Therefore, the $\mathbf{r}$-dependent part of the wave function is $\Psi_2(\mathbf{r})=\psi_X(r)e^{im\varphi}$, where $\varphi$ is the polar angle in the $x,y$ plane. In what follows, we will only consider the exciton states with $m=0$ since they can be directly optically probed (``bright states''). As a result, we arrive from the Hamiltonian (\ref{Hefff}) at the one-dimensional Schr\"odinger equation,
\begin{equation}\label{Sr}
\left[-\frac{\hbar^2}{2\mu}\frac{1}{r}\frac{\partial}{\partial r}\left(r\frac{\partial}{\partial r}\right)+\frac{\mu\Omega^2{r}^2}{2}+U_0({r})\right]\psi_X(r)
={\cal E}\psi_X(r),
\end{equation}
which defines both the exciton binding energy, $\varepsilon_X=-{\cal E}$, and the corresponding wave function, $\psi_X(r)$, for the exciton states with the zero angular momentum.

For definiteness, let us consider a GaAs-based QD with the permittivity $\epsilon\approx 12$ and the effective masses of electrons and holes $m_e=0.067m_0$ and $m_h=0.47m_0$, respectively, where $m_0$ is the mass of electron in vacuum. The typical size of such a QD ranges from several nanometers to few tens of nanometers. Therefore, we will restrict the consideration by the two limiting cases: A small QD with the effective size $L=3$~nm and a large QD with  $L=15$~nm. Taking into account
the relationship between the confining potential frequency, $\Omega$, and the effective QD size~\cite{que1992excitons} $$L=\sqrt{\frac{\hbar}{\mu\Omega}},$$ one can solve Eq.~(\ref{Sr}) numerically for the above-mentioned QD with using the standard numerical shooting method~\cite{indjin1995numerical}. The calculation results are presented below in Figs.~2 and 3 for the ground exciton state and different irradiation intensities $I=cE_0^2/4\pi$.

\begin{figure}[h!]
\centering\includegraphics[width=1.0\columnwidth]{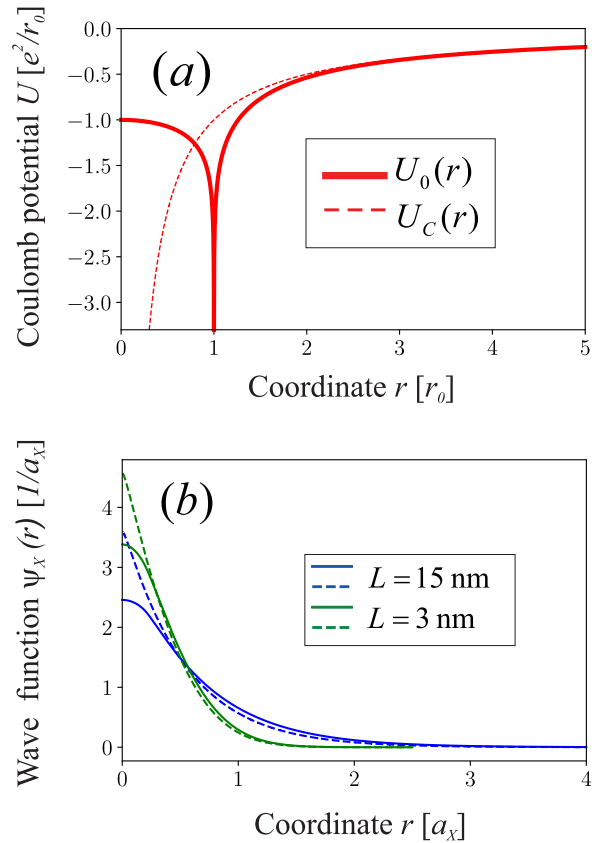}
\caption{(a) Coordinate dependence of the dressed Coulomb potential $U_0(r)$ (solid line) and the bare Coulomb potential $U_C(r)$ (dashed line); (b) Coordinate dependence of the exciton wave function $\psi_X(r)$ of the ground exciton state for QDs with the effective size $L=3$~nm and $L=15$~nm in the presence of the irradiation with the photon energy $\hbar\omega_0=1$~meV and the intensity $I=20$~mW/cm$^2$ (solid lines) and in the absence of the irradiation (dashed lines), where $a_X=\hbar^2\epsilon/\mu e^2$ is the exciton Bohr radius.}\label{fig2}
\end{figure}
The spatial profile of both the dressed Coulomb potential (\ref{U00}) and the bare Coulomb potential (\ref{UC}) are plotted in Fig.~2a. Comparing the dressed Coulomb potential (the solid line) and the bare Coulomb potential (the dashed line), one can conclude that the dressing field (\ref{A}) induces the repulsive area near $r=0$, whereas the attractive Coulomb well is shifted by the field from the point $r=0$ to the ring of the radius $r=r_0$. This repulsive area increases the effective distance between electron and hole and, correspondingly, decreases localization of the exciton wave function $\psi_X(r)$ (see Fig.~2b). As a consequence, the exciton binding energy $\varepsilon_X$ also decreases (see Fig.~3a).
\begin{figure}[h!]
\centering\includegraphics[width=0.8\columnwidth]{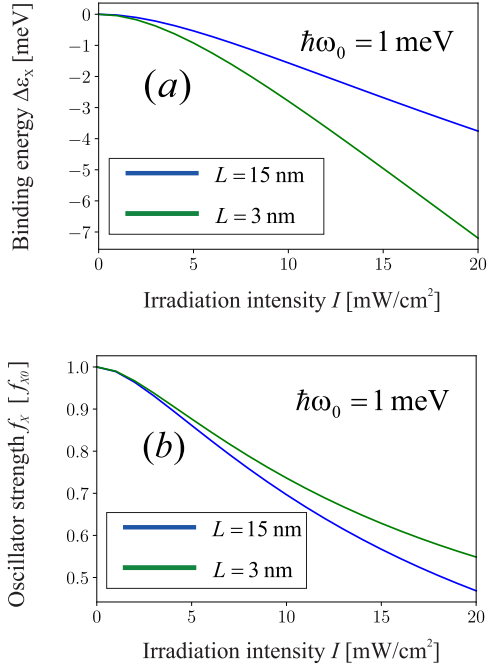}
\caption{(a) Dependence of the field-induced shift of the exciton binding energy, $\Delta\varepsilon_X=\varepsilon_X-\varepsilon_{X0}$, for the ground exciton state on the irradiation with the photon energy $\hbar\omega_0=1$~meV and intensity $I$ for QDs with the effective size $L=3$~nm and $L=15$~nm, where $\varepsilon_{X0}$ is the exciton binding energy in the absence of irradiation; (b) Dependence of the exciton oscillator strength, $f_X$, on the irradiation with the photon energy $\hbar\omega_0=1$~meV and intensity $I$ for the same exciton state and QDs, where $f_{X0}$ is the oscillator strength in the absence of irradiation.}
\label{fig3}
\end{figure}

The dependence of the exciton oscillator strength, $f_X$, on the irradiation intensity, $I$, is plotted in Fig.~3b. In semiconductor QDs, the exciton oscillator strength reads~\cite{que1992excitons}
\begin{equation}\label{f}
f_X=\frac{2|p_{cv}|^2}{m_0(\varepsilon_g+\bar{\varepsilon}_X)}|\psi_X(0)|^2\left|\int_S \Psi_1(\mathbf{R})d^2\mathbf{R}\right|^2,
\end{equation}
where $\bar{\varepsilon}_X=V_0+\hbar\Omega/2-\varepsilon_X$ is the total exciton energy, $p_{cv}$ is the interband momentum matrix element in the semiconductor, $\varepsilon_g$ is the semiconductor band gap, and the integration area $S$ is the $x,y$ plane as a whole. The oscillator strength (\ref{f}) defines the radiative broadening of the exciton linewidth~\cite{citrin1993radiative},
\begin{equation}\label{G}
\Gamma_X= \frac{\hbar e^2}{m_0L^2c\sqrt{\epsilon}}f_X,
\end{equation}
and the corresponding exciton lifetime, $\tau_X=\hbar/2\Gamma_X$. It follows from Eq.~(\ref{f}) that the oscillator strength depends on the wave function $\psi_X(r)$ at $r=0$.
Since the irradiation induces the repulsive area near $r=0$ (see Fig.~2a), it decreases the wave function $|\psi_X(0)|$ (see Fig.~2b). As a consequence, the irradiation also decreases the oscillator strength $f_X$ (see Fig.~3b) and, correspondingly, increases the exciton lifetime $\tau_X$. Thus, the stabilization of the exciton by the high-frequency field (dynamical stabilization) appears. It follows from Fig.~3 that the exciton binding energy $\varepsilon_X$, the oscillator strength (\ref{f}) and the radiative broadening (\ref{G}) can be substantially decreased by the relatively weak irradiation. This opens the way for the all-optical control of both the exciton binding energy and the exciton radiative lifetime of the QD-based single photon emitters, which could find its applications in the quantum optical communication setups. Since the onset of field-induced effects happens at the timescale of the field period, the discussed method to control excitonic properties of QDs by a high-frequency electromagnetic field is very fast as compared to the relatively slow electrostatic control of them by the gate voltage~\cite{hallett2018electrical, trivedi2020generation}.

The effects discussed above --— the decreasing of exciton binding energy and oscillator strength with increasing field —-- are originated physically from the field-induced repulsive area in the electron-hole interaction potential near $r=0$ (see Fig. 2a). Therefore, these effects depend mainly on the electron-hole interaction and remain the same qualitatively for any realistic confinement. It should be noted that the considered parabolic confinement potential (\ref{UOm}) with the same frequency $\Omega$ for both electron and hole is the model which was introduced into the quantum dot theory in order to separate the centre of mass coordinate of the electron-hole pair $\mathbf{R}$ and its relative coordinate $\mathbf{r}$ in the exciton Hamiltonian~\cite{que1992excitons}. Since such a separation of variables substantially simplifies the analysis of excitonic effects, this conventional model was applied above to describe the field-induced effects in the simplest way. It should be noted also that an oscillating field does not change the distance between identical charged particles and, therefore, does not modify interaction of them~\cite{Kibis_2019}. Thus, only the electron-hole interaction is altered by the electromagnetic field, whereas the electron-electron and hole-hole interactions remain unaffected. As a consequence, the discussed field-induced effects are expected to be the same qualitatively for trions~\cite{Tischler_2002} in charged quantum dots and biexcitons~\cite{Yoffe_2001}. Concerning applicability limits of the developed theory, the present analysis is correct if the exciton lifetime, $\tau_X$, is far larger than the dressing field period, $T=2\pi/\omega_0$. As a consequence, the condition
\begin{equation}\label{tau}
\omega_0\tau_X\gg1
\end{equation}
should be satisfied. In state-of-the-art semiconductor QDs, the lifetime $\tau_X$ is of nanosecond scale and, therefore, the developed theory is applicable for dressing field frequencies $\omega_0$ starting from the microwave range.

It follows from the aforesaid that the above-discussed excitonic effects appear due to the crucial change of the dressed Coulomb potential (\ref{U00}) for small distances $r$, where the dressing field (\ref{A}) induces the repulsive area (see the solid line in Fig.~2a). It should be noted that the inverse physical situation takes place for the repulsive Coulomb interaction. In this case, the circularly polarized field (\ref{A}) induces the attractive area in the core of the repulsive Coulomb potential, which can lead to the electron states bound at various repulsive potentials and, particularly, to the light-induced electron pairing~\cite{Kibis_2019,Kibis_2020_2,Kibis_2021_2,Kibis_2021_3,Iorsh_2021}.

\section{Conclusion}
We have demonstrated that the electromagnetic irradiation of relatively weak intensity allows for the dynamical control over the binding energy and radiative lifetime of excitons in semiconductor quantum dots (QDs). The effect originates from the renormalization of the electron-hole attractive Coulomb potential by the field which induces the repulsive area in the core of the attractive potential. This method allows for the ultrafast control over the exciton spectral characteristics, which can find its application in QD-based platforms for optical quantum communications and quantum metrology.

\begin{acknowledgments}
The reported study was funded by the Russian Science Foundation
(project 20-12-00001).
\end{acknowledgments}

\appendix
\section{The Floquet problem}
In the most general form, the nonstationary Schr\"odinger equation
for an electron (a hole) in a periodically time-dependent field with the
frequency $\omega_0$ can be written as
$i\hbar\partial_t\psi(t)=\hat{\cal H}(t)\psi(t)$, where $\hat{\cal
H}(t+T)=\hat{\cal H}(t)$ is the periodically time-dependent
Hamiltonian and $T=2\pi/\omega_0$ is the field period. It follows
from the well-known Floquet theorem that solution of the Schr\"odinger
equation is the Floquet function, $\psi(t)=e^{-i\varepsilon
t/\hbar}\varphi(t)$, where $\varphi(t+T)=\varphi(t)$ is the
periodically time-dependent function and $\varepsilon$ is the
electron (quasi)energy describing behavior of the electron in the
periodical field. The Floquet problem is aimed to find the electron energy spectrum, $\varepsilon$. To solve the problem, let us introduce the unitary transformation, $\hat{\cal U}=e^{iS}$, which transfers the time dependence from the Hamiltonian $\hat{\cal H}(t)$ to its basis states. Then we arrive from the time-dependent Hamiltonian $\hat{\cal H}(t)$  to the effective time-independent Hamiltonian
\begin{equation}\label{HefA}
\hat{\cal H}_{\mathrm{eff}}=\hat{\cal U}^\dagger\hat{\cal H}\hat{\cal U} -
i\hbar\hat{\cal U}^\dagger\partial_t
\hat{\cal U}.
\end{equation}
Solving the stationary Schr\"odinger problem with the Hamiltonian (\ref{HefA}), $\hat{\cal H}_{\mathrm{eff}}\Psi=\varepsilon\Psi$, one can find the sought electron energy spectrum, $\varepsilon$.

There is the regular method to find the transformation matrix $S$  as the $1/\omega_0$ expansion (the Floquet-Magnus expansion)~\cite{Eckardt_2015,Goldman_2014,Bukov_2015,Casas_2001}. Keeping the first three terms in the expansion, the effective stationary Hamiltonian (\ref{HefA}) reads
\begin{align}\label{HeA}
&\hat{\cal H}_{\mathrm{eff}}=\hat{\cal H}_{0}+\sum_{n=1}^\infty\frac{[\hat{\cal H}_{n},\hat{\cal H}_{-n}]}{n\hbar\omega_0}\nonumber\\
&+\frac{1}{2(\hbar\omega_0)^2}\sum_{n=1}^\infty\frac{1}{n^2}\Big([[\hat{\mathcal{H}}_n,\hat{\mathcal{H}}_0],\hat{\mathcal{H}}_{-n}]+\mathrm{H.c.}\Big),
\end{align}
where $\hat{\cal H}_n$ are the harmonics of the Fourier expansion  $\hat{\cal H}=\sum_{n=-\infty}^{\infty}\hat{\cal H}_ne^{in\omega_0 t}$ . Since the second term of the Floquet-Magnus expansion (\ref{HeA}) for the Hamiltonian (\ref{Ht}) is zero, the addition to the main Hamiltonian (\ref{Hef}) is defined by the last term of the Hamiltonian (\ref{HeA}). To describe the harmonics arisen from the Coulomb potential $U_C(\mathbf{r}_e-\mathbf{r}_{e}^\prime(t),\mathbf{r}_h-\mathbf{r}_{h}^\prime(t))$ in the Hamiltonian (\ref{Ht}), let us use the course-of-value function,
\begin{equation}\label{Pl}
\sum_{m=0}^{\infty} P_m(z)x^m=\frac{1}{\sqrt{1-2xz+x^2}},
\end{equation}
where $$P_m(z)=\frac{1}{2^mm!}\frac{d^m}{x^m}(z^2-1)^m$$
is the Legendre polynomial~\cite{GR_book}.
Applying Eq.~(\ref{Pl}) and omitting the coordinate-independent terms, the effective Hamiltonian (\ref{HeA}) can be written as
\begin{equation}\label{HA}
\hat{\cal H}_{\mathrm{eff}}=\frac{\hat{\mathbf{P}}^2}{2M}+\frac{M\Omega^2\mathbf{R}^2}{2}+\frac{\hat{\mathbf{p}}^2}{2\mu}+\frac{\mu\Omega^2{r}^2}{2}+U_0({r})+V(r),
\end{equation}
where
\begin{equation}\label{V}
V(r)=\frac{2}{\mu\omega_0^2}\sum_{n=1}^\infty\left(\frac{\partial V_n}{\partial r}\right)^2,
\end{equation}
is the potential arisen from the last term of the Floquet-Magnus expansion (\ref{HeA}), and
\begin{align}\label{VnA}
&V_n({r})=-\left(\frac{e^2}{2\pi\epsilon}\right)\nonumber\\
&\times\left\{\begin{array}{rl}
\frac{1}{r_0}\sum_{m=0}^\infty\left(\frac{r}{r_0}\right)^m\int_0^{2\pi} P_m(\cos\theta)e^{in\theta} d\theta,
&\frac{r}{r_0}<1\\\\
\frac{1}{r}\sum_{m=0}^\infty\left(\frac{r_0}{r}\right)^m\int_0^{2\pi} P_m(\cos\theta)e^{in\theta} d\theta,
&\frac{r}{r_0}>1
\end{array}\right.
\end{align}
are the Fourier harmonics of the Coulomb potential $U_C(\mathbf{r}_e-\mathbf{r}_{e}^\prime(t),\mathbf{r}_h-\mathbf{r}_{h}^\prime(t))$.
It should be reminded that the effects discussed in the present article appear due to the field-induced local maximum of the potential $U_0(r)$ near $r=0$ (see Fig.~2a). Comparing Eq.~(\ref{Hefff}) and Eq.~(\ref{HA}), one can conclude that the approximation of the effective Hamiltonian (\ref{HeA}) by the main term (\ref{Hef}) is correct to describe these effects if the contribution of the potential $U_0(r)$ to the Hamiltonian (\ref{HA}) much exceeds the contribution of the potential $V(r)$ for $r\ll r_0$. As a result, we arrive at the applicability condition of the developed Floquet theory,
\begin{equation}\label{CA}
\frac{e^2}{\epsilon\mu r_0^3\omega_0^2}\ll1,
\end{equation}
which can be satisfied for varied irradiation intensities within the broad frequency range defined by Eq.~(\ref{tau}).

\end{document}